\documentclass[reprint, amsmath,amssymb,showpacs,floatfix, aps, prl, twocolumn, superscriptaddress]{revtex4-1}
\usepackage{graphicx}
\usepackage{dcolumn}
\usepackage{bm}
\usepackage{color}
\usepackage{graphicx}
\usepackage{gensymb}
\usepackage{braket}
\usepackage{hyperref}
\hypersetup{colorlinks=true,citecolor=blue,linkcolor=blue, urlcolor=blue}
\def\nn{\nonumber}
\def\({\left(}
\def\){\right)}
\def\[{\left[}
\def\]{\right]}

\def\e{\epsilon}
\def\l{\lambda}

\def\hb{\hbar}

\def\bfA{\mathbf{A}}
\def\bfQ{\mathbf{Q}}

\def\bfe{\mathbf{e}}
\def\bfk{\mathbf{k}}
\def\bfp{\mathbf{p}}
\def\bfq{\mathbf{q}}

\def\>{\rangle}
\def\<{\langle}

\begin{document}
\title{Theory and \textit{Ab Initio} Computation of the Anisotropic Light Emission\\ in Monolayer Transition Metal Dichalcogenides}
\author{Hsiao-Yi Chen}
\affiliation {Department of Applied Physics and Materials Science, Steele Laboratory, California Institute of Technology, Pasadena, California 91125, United States}
\affiliation {Department of Physics, 
California Institute of Technology, Pasadena, California 91125, United States}
\author{Maurizia Palummo}
\affiliation {Dipartimento di Fisica and INFN, Universit$\grave{a}$  di Roma Tor Vergata€ Via della Ricerca Scientifica 1, 00133 Roma, Italy}
\author{ Davide Sangalli}
\affiliation {CNR-ISM, Division of Ultrafast Processes in Materials (FLASHit), Area della Ricerca di Roma 1, Monterotondo Scalo, Italy}
\author{Marco Bernardi}
\email{bmarco@caltech.edu}
\affiliation {Department of Applied Physics and Materials Science, Steele Laboratory, California Institute of Technology, Pasadena, California 91125, United States}

\begin{abstract}
\noindent
Monolayer transition metal dichalcogenides (TMDCs) are direct gap semiconductors with unique potential for ultrathin light emitters. Yet, their photoluminescence (PL) is not completely understood. 
We compute the radiative recombination rate in monolayer TMDCs as a function of photon emission direction and polarization, and obtain polar plots of the PL for different excitation scenarios  
using the \textit{ab initio} Bethe-Salpeter equation.  
We show that excitons in a quantum superposition state of the K and K' inequivalent valleys emit light anisotropically upon recombination. Our results can explain the PL anisotropy and polarization dependence measured in recent experiments, and predict new light emission regimes. 
When averaged over emission angle and exciton momentum, our new treatment recovers the temperature dependent radiative lifetimes we previously derived. 
Our work provides a first-principles approach to study light emission in two-dimensional materials.
\end{abstract}
%\pacs{--} 
\maketitle

%
%---------   MAIN TEXT     -----------%
%  
Two-dimensional transition metal dichalcogenides (2D-TMDCs) with chemical formula $\rm MX_2$ (M=Mo, W and X=S, Se, Te) are lead candidates for novel optoelectronic devices 
\cite{Strano,Maier,Bernardi-review,Bernardi-ML,Britnell,Atwater,Splendiani,Heinz,Palummo,Amani}.
They exhibit a direct gap in their monolayer form and an indirect gap in bulk crystals and multi-layers.
Monolayer TMDCs can absorb light strongly \cite{Bernardi-ML}, and due to their direct gap are expected to also emit light efficiently. However, experiments on exfoliated monolayers typically 
exhibit weak photoluminescence (PL) \cite{Heinz}. Recent work reported near-unity PL quantum yield in $\rm MoS_2$ \cite{Amani}, but its origin is still debated \cite{Tisdale}. 
While their radiative recombination has been investigated using time-resolved spectroscopy \cite{Shi,Korn,Lagarde,Libai} and \textit{ab initio} calculations \cite{Palummo}, 
microscopic understanding of light emission in 2D-TMDCs remains incomplete.\\
\indent
The lack of inversion symmetry in monolayer TMDCs leads to two inequivalent valleys at the K and K' corners of the hexagonal Brillouin zone. Locking of the spin and valley degrees of freedom introduces optical valley selection rules \cite{Xiao-2,Cao,Xu-spin}, whereby circularly polarized light can be employed to selectively generate excitons in a specific valley \cite{Xu-spin,Zeng, Mak-valley,Xiao}. 
As a result, linearly polarized light can form excitons in a quantum superposition of the two valleys, and linearly polarized PL can probe the coherence of such excitonic states \cite{Jones, Wang, Ye-Heinz}.\\ 
\indent
An important result that has received limited attention is that the linearly polarized PL seen experimentally is anisotropic \cite{Jones,Wang, Ye-Heinz} in spite of the in-plane isotropic hexagonal structure of 2D-TMDCs. 
The intensity of this anisotropic PL has also been seen to depend strongly on light polarization \cite{Jones}. 
Theory and experiments have also shed light on valley decoherence \cite{Yu-valley,Yu-Hall,Alejandro,Jones,Mai,Hao}, but quantifying exciton coherence through the PL remains an open problem. 
Understanding exciton dynamics, decoherence and light emission is critical to advancing 2D-TMDCs.\\
\indent
%
%------   HERE WE SHOW    ----%
%
Here, we derive and compute the radiative rates as a function of photon emission direction and polarization in monolayer TMDCs.  
We employ the \textit{ab initio} Bethe-Salpeter equation (BSE) to compute exciton energies and wavefunctions \cite{Palummo}. 
The lowest-energy eigenvectors of the BSE are rotated in their degenerate subspace to form excitons with different valley superposition states. 
Polar plots of the PL generated when these excitons recombine can explain recent PL measurements under excitation with linearly polarized light, and predict new light emission regimes. 
Our approach is general, and it enables \textit{ab initio} calculations of the PL in 2D semiconductors. 
Our results shed light on the physics of light emission in 2D-TMDCs, explaining their PL anisotropy and its link to valley polarization and decoherence.\\
\indent
% 
%%%      COMPUTATIONAL METHODS
% 
We carry out density functional theory (DFT) calculations within the generalized gradient approximation \cite{PBE} using the {\sc Quantum Espresso} code \cite{QE}. Experimental lattice parameters are used, together with fully relativistic pseudopotentials that include the spin-orbit coupling and treat semi-core states as valence electrons \cite{Palummo,Lambrecht}. The Yambo code \cite{Yambo} is employed to solve the BSE using a $33\times 33\times 1$ $\mathbf{k}$-point grid. A rigid shift of the conduction band DFT eigenvalues is applied to obtain quasiparticle bandstructures consistent with GW \cite{Palummo}.\\ 
\indent
%
%%%       THEORY
%
Within the Tamm-Damcoff approximation, an exciton in state $S$ with center-of-mass momentum $\bfQ$ can be written as a coherent superposition of electron-hole pairs:
\begin{equation}
\label{Tamm-Damcoff}
|S\bfQ\>=\sum_{vc\bfk}A^{S\bfQ}_{vc\bfk}|v\bfk\>|c\bfk+\bfQ\>
\end{equation} 
where $v$ and $c$ label the valence and conduction bands, $\bfk$ is the electron crystal momentum, and the coefficients $A^{S\bfQ}_{vc\bfk}$ are obtained by solving the BSE. 
The interaction between electrons and photons is treated using the Hamiltonian $H^{\textrm{int}} \!=\! \frac{e}{m} \bfA\cdot \bfp$, where $\bfp$ is momentum and $\bfA$ the vector potential in second quantized form \cite{Loudon}.
Following our previous work \cite{Palummo}, we employ Fermi's golden rule to obtain the exciton radiative decay rate: 
\begin{eqnarray}
\label{FGR}
&&\gamma_S(\bfQ)=\frac{2\pi}{\hbar}
\sum_{\l\bfq} \left|\langle G, 1_{\l \bfq}|H^{\rm int}|S\bfQ, 0\rangle\right|^2 \delta(E_S(\bfQ)-\hbar cq)\nn\\
&&~~~~~=
\frac{\pi e^2}{\e_0 m^2cV} \sum_{\l\bfq} \frac{1}{q} \left|\bfe_{\l\bfq } \cdot \bfp_{S}(\bfQ)\right|^2 \delta(E_S(\bfQ)-\hbar cq)\nn\\
\end{eqnarray} 
where the initial state $|S\bfQ,0\>$ is an exciton with no photon, and the final state $ |G, 1_{\lambda \bfq}\>$ the ground state with one emitted photon. %,  
%
% FIGURE 1 --- COORDINATES
%
\begin{figure}[!t]
\centering
\includegraphics[scale=0.28]{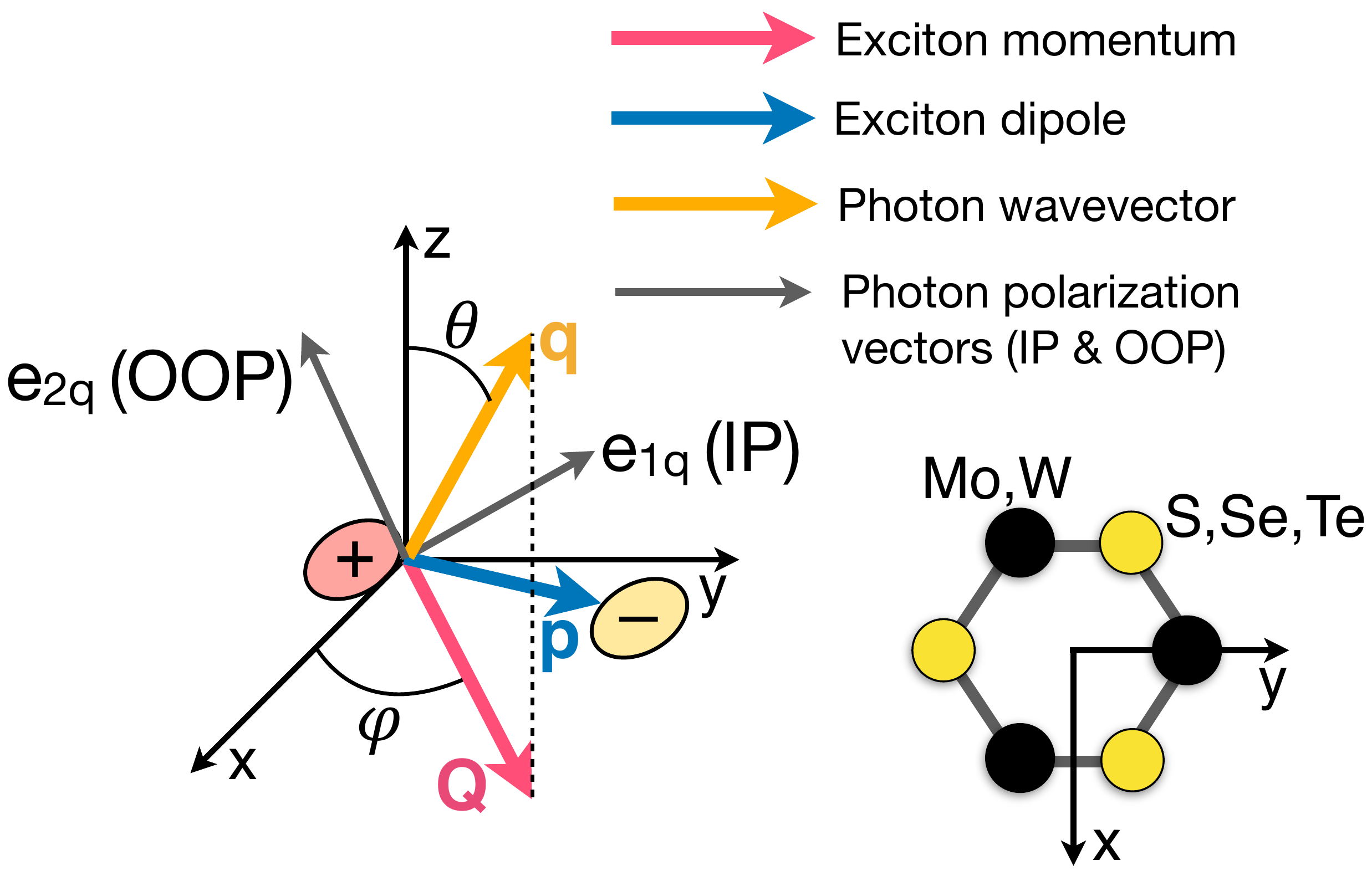}
\caption{Physical quantities entering our equations. The exciton dipole and center-of-mass momentum are shown, together with the photon wavevector and the in-plane (IP) and out-of-plane (OOP) polarization vectors, $\bfe_{1\bfq}$ and $\bfe_{2\bfq}$ respectively. The inset shows the cartesian coordinates relative to the crystal structure, where spheres represent the atoms.}
\label{Fig:coordinate}
\end{figure} 
The sum runs over two polarizations $\lambda = 1,2$ (with polarization vectors $\mathbf{e}_{\lambda \bfq}$) and the wavevector $\mathbf{q}$ of the emitted photon.  
Since we focus on monolayers, the exciton center-of-mass momentum is a vector $\bfQ=Q_x \hat{\bf x} + Q_y \hat{\bf y}$ in the $xy$ plane containing the material. Momentum conservation thus requires the in-plane component of the emitted photon wavevector to be equal to $\bfQ$, namely, $\bfq = \bfQ + q_z \hat{\bf z}$ (see Fig.~\ref{Fig:coordinate}).\\
\indent
%
% STANDARD BSE, SMALL Q, DIPOLE DEFINITION
% 
The transition dipole in Eq.~\ref{FGR}, $\bfp_{S}(\bfQ) \!=\! \langle G|\mathbf{p}|S\bfQ\rangle$ 
\footnote{In practice, we use the velocity operator, and compute the transition dipole as $\mathbf{p}_S (\mathbf{Q})\!=\! (-i m / \hbar) \braket{G| \[ \mathbf{x},H \]| S\mathbf{Q}}$ to correctly include the non-local part of the Hamiltonian \cite{Davide}.}, is called hereafter the dipole of exciton $S$. 
Since 2D materials have a weak optical response in the layer-normal direction, we can ignore the $z$-component of the dipole. 
For light emission, the values of $Q$ compatible with energy conservation are very small. For this reason, we approximate the dipole of exciton $\ket{S\bfQ}$ as $\bfp_S(\bfQ)\!\approx\! \bfp_S(0)$ by solving the BSE at $\bfQ\!=\!0$ (the BSE with finite $\bfQ$ \cite{Gatti-BSE} has been solved for 2D-TMDCs in Ref. \cite{Diana}).   
Note that the components of $\mathbf{p}_S$ are in general \textit{complex numbers}. We previously treated the special case in which $\mathbf{p}_S$ is real and arbitrarily chosen to be in the $x\,$=$\,y$ direction \cite{Palummo}.  
This work generalizes the result to an arbitrary complex $\bfp_S$, leading to rich physical consequences.\\
\indent
%
% COORDINATE DESCRIPTION
%
Using the coordinates in {Fig.~\ref{Fig:coordinate}}, we write the transition dipole as $\mathbf {p}_S=p_{Sx}\,\hat{\bf x}+p_{Sy}\,\hat{\bf y}$, with complex $p_{Sx}$ and $p_{Sy}$. 
Without loss of generality, the polarization vectors $\bfe_{\lambda \mathbf{q}}$ of the emitted photon are chosen as the in-plane (IP) and out-of-plane (OOP) unit vectors   
\footnote{The IP and OOP polarizations are also referred to in the literature as the horizontal and vertical polarizations, respectively, or the transverse (IP) and longitudinal (OOP) polarizations in Ref.~\cite{Diana}}: 
\begin{eqnarray}
\label{Eq photo polar}
{\rm IP:~}&& \bfe_{1\bfq}
=(-\sin\varphi, \cos\varphi,0)\nonumber\\
{\rm OOP:~}&& \bfe_{2\bfq}
=(-\cos\theta\cos\varphi, -\cos\theta\sin\varphi,\sin\theta),
\end{eqnarray}
where $\varphi$ is the angle between the $x$-axis and $\bfQ$ (and thus between the $x$-axis and the in-plane projection of $\bf q$).  
%
% MAIN RESULT, EQ. 4
%
For an exciton with momentum $\bfQ$, the total radiative rate is obtained by summing over both polarizations in Eq.~(\ref{FGR}). 
We obtain (see the Supplemental Material \cite{Supp}):
\begin{widetext}
\begin{equation}
\label{gamma0}
\gamma_S(\bfQ) \!=\! \gamma_S(0)\!\cdot\!
\left(\frac{E_S(0)}{\sqrt{E_S^2(Q)-\hbar^2c^2Q^2}}\right) \left\{
\left|- \frac{ p_{Sx}}{p_S}\sin\varphi+\frac{ p_{Sy}}{p_S}\cos\varphi \right|^2_{\rm IP} + 
\frac{E_S(Q)^2-\hbar^2c^2Q^2}{E_S(Q)^2}\left|\frac{ p_{Sx}}{p_S}\cos\varphi +\frac{ p_{Sy}}{p_S}\sin\varphi \right|_{\rm OOP}^2
\right\}
\end{equation}
\end{widetext}
where $E_S(0)$ is the exciton energy computed with the BSE, $E_S(Q)$ the finite-momentum exciton energy, and $\gamma_S(0)=\frac{e^2p_S^2}{\epsilon_0m^2cA E_S(0)}$ the radiative rate for $\bfQ\!=\!0$; the two terms in curly brackets correspond, respectively, to the IP and OOP emitted photon polarizations. 
Due to momentum conservation, there is an upper value of $Q\!=\!Q_0$ for radiative decay, given by the light-cone condition $E_S(Q_0)=\hb c Q_0$;  
the radiative rate vanishes for $Q\!>\!Q_0$.\\
\indent
%
% FIGURE 2 --- POLAR PLOTS OF THE RADIATIVE RATES
%
\begin{figure*}[t]
\centering
\includegraphics[scale=.4]{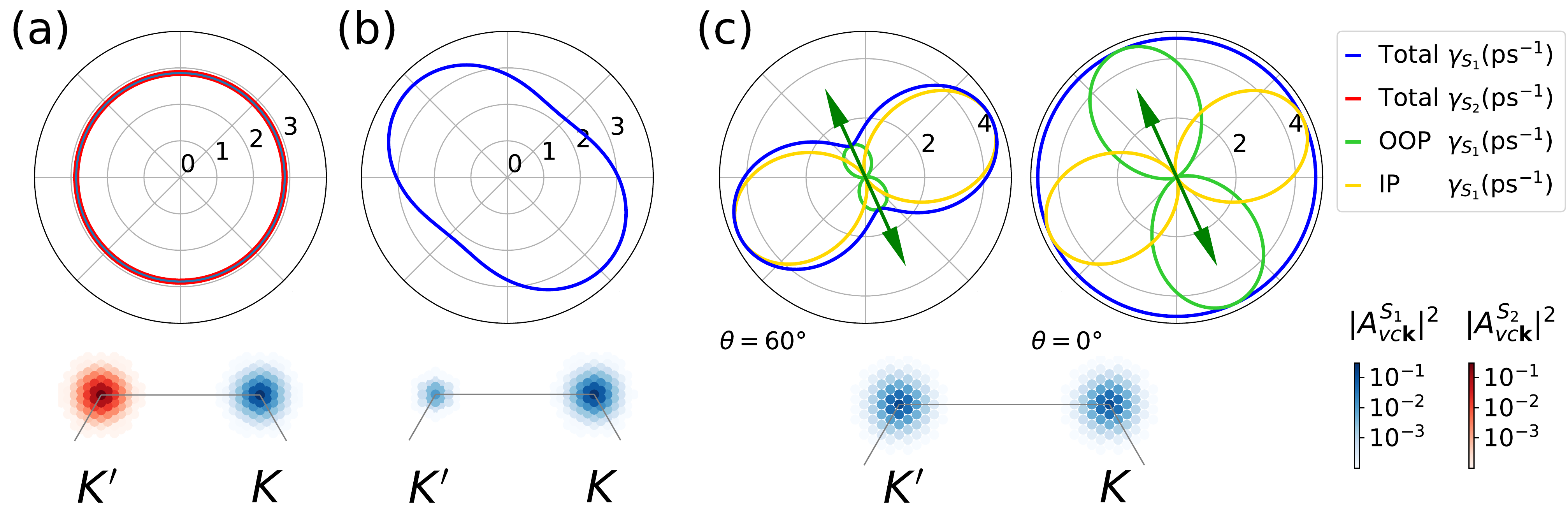}
\caption{Polar plots of the radiative rates, and the corresponding exciton wavefunctions, shown for several cases. (a) Two distinct excitons entirely located, respectively, on the K and K' valleys, and their isotropic radiative rate. (b) Exciton with unequal weights on the K and K' valleys, and the resulting anisotropic radiative rate and PL emitted for $\theta=60\degree$.  (c) Exciton with equal weights on the K and K' valleys, as generated by linearly polarized light, and its radiative rate emitted at a polar angle $\theta=60\degree$ (left panel) and along the layer normal at $\theta=0\degree$ (right panel). The rates for OOP and IP polarized light emission are shown along with their sum. The arrow shows the polarization direction of incident light.}
\label{fig2}
\end{figure*}
%
% COSINE FACTOR AND POLARIZATION DEPENDENT FORMULAS  GAMMA(THETA,PHI)
%
We compute the dependence of the radiative rate on the polar angle $\theta$ between the photon emission direction and the layer normal (see Fig.~\ref{Fig:coordinate}). 
Using $E_S(Q)\!\approx\! E_S(0)$ due to the very small exciton momentum inside the light cone, together with simple geometric arguments, we have:
\begin{equation}
\frac{\sqrt{E_S^2(Q)-\hbar^2c^2Q^2}}{E_S(0)}\approx
\frac{\sqrt{E_S^2(Q)-\hbar^2c^2Q^2}}{E_S(Q)}=\cos\theta.
\end{equation}
Substituting in Eq.~\ref{gamma0}, and using $\gamma_S(\theta,\phi) = \gamma_S(\mathbf{Q}) \cos(\theta)$ \cite{Supp}, we obtain the radiative rates for light emitted with IP and OOP polarizations:
\begin{align}
\label{IPOOP}
&\gamma_S^{\rm IP}(\theta,\varphi) \,\,\,\,=\,\, \gamma_S(0) \, \left|-\frac{ p_{Sx}}{p_S}\sin\varphi +\frac{ p_{Sy}}{p_S}\cos\varphi \right|^2 \\
&\gamma_S^{\rm OOP}(\theta,\varphi) \!= \gamma_S(0) \cos^2\theta  \left|\frac{ p_{Sx}}{p_S}\cos\varphi +\frac{ p_{Sy}}{p_S}\sin\varphi \right|^2.
\end{align}
Since the intensity of light emitted at a given angle is proportional to the radiative rate, these equations can provide polar plots of the PL. 
The IP and OOP contributions, which can be measured separately in experiments able to discern the PL polarization, can be added together to obtain the total PL intensity.\\
\indent 
% 
% EXCITON DEGENERACY AND ROTATION
%
An important point is that the lowest-energy exciton responsible for light emission (so-called bright $A$ 1$s$ exciton \cite{Palummo, Heinz})
is two-fold degenerate in 2D-TMDCs due to the valley degeneracy. 
These degenerate excitons, called here $|S_1\>$ and $|S_2\>$, are orthogonal but randomly oriented in their degenerate subspace when the BSE Hamiltonian is diagonalized numerically at $\bfQ\!=\!0$. 
They can be rotated in the degenerate subspace to new states $\ket{S'_i} = {\bf M}_{ij} \ket{S_j}$ using a unitary matrix {\bf M} in SU(2) \cite{Lie}:
\begin{equation}
{\bf M} (u,\theta_1,\theta_2)
=
\begin{pmatrix} u & \sqrt{1-|u|^2}e^{i\theta_1} \\ -\sqrt{1-|u|^2}e^{ -i(\theta_1-\theta_2)  } & u^{ * }e^{ i\theta_2  } \end{pmatrix}
\end{equation} 
where $u$, $\theta_1$ and $\theta_2$ are independent parameters defining the transformation. 
Since excitons are represented by coefficients $A_{vc\bfk}^{S}$ in the electron-hole basis employed to solve the BSE \cite{Yambo,Rohlfing}, the rotation is accomplished by transforming 
the exciton coefficients as $( A_{vc\bfk}^{S'_1},A_{vc\bfk}^{S'_2})^T = {\bf M} \cdot(A_{vc\bfk}^{S_1},A_{vc\bfk}^{S_2})^T$, where T is the transpose.\\
\indent 
In the following, the transformed excitons $\ket{S'_i}$ are chosen as those physically relevant in selected excitation scenarios of interest. 
The square modulus of their coefficients, $|A_{vc\bfk}^{S'_i}|^2$, define the probability to find the exciton in the K and K' valleys. 
The exciton dipoles, by virtue of their definition $\bfp_{S_i}\!=\! \braket{G|\mathbf{p}|S_i}$, transform in the same way as the exciton states, 
namely $\mathbf{p'}_{\!i} \equiv \mathbf{p}_{S'_i} \!=\! \mathbf{M}_{ij} \mathbf{p}_{S_j}$.
The dipoles $\bfp'_{1,2}$ of the transformed excitons determine their radiative rate through Eq.~\ref{gamma0}.\\  
\indent
%
% FIGURE 2
%
Figure \ref{fig2} shows different excitation and light emission scenarios. For each case, we plot the exciton weights $\left| A^{S}_{vc \bfk}\right|^2$ on the two valleys and the radiative rate $-$ which is proportional to the intensity of the PL signal $-$ as a function of in-plane light emission angle $\varphi$ at a fixed polar angle $\theta$. The results shown here are for WSe$_2$, but similar trends also hold for other 2D-TMDCs.\\
\indent 
Fig.~\ref{fig2}(a) focuses on excitons generated with circularly polarized light. We transform the BSE eigenvectors to obtain two excitons $\ket{S_{1,2}}$ each located entirely on one valley. We find that the PL for these excitons is isotropic about the layer normal, regardless of the angle $\theta$ at which light emission is detected. The isotropic PL is consistent with the fact that circularly polarized photons cannot break the in-plane rotational symmetry of 2D-TMDCs.\\ 
\indent
In Fig.~\ref{fig2}(b), we form excitons with unequal weights on the K and K' valleys, which can be directly excited with light or result from decoherence processes.  
By placing more weight on either valley, the isotropic PL pattern is broken $-$ the radiative rate becomes greater along a specific direction, and the PL is anisotropic.\\ %polarizations other than circular or linear. 
\indent
%
% FIG 2C, LINEARLY POLARIZED LIGHT
%  
Figure~\ref{fig2}(c) focuses on excitons generated with incident light linearly polarized in the $\hat{\mathbf{E}}_{\textrm{inc}}$ direction.   
We form two excitons $|S_{1,2}\>$ with, respectively, dipoles $\bfp_1$ parallel and $\bfp_2$ perpendicular to $\hat{\mathbf{E}}_{\textrm{inc}}$. 
With this choice, only $|S_1\>$ is excited since $|\bfp_2 \cdot \hat{\mathbf{E}}_{\textrm{inc}}| = 0$. 
Consistent with the optical valley rule, the resulting exciton $|S_1\>$ is an equal superposition state of the K and K' valleys, further proving the validity of our rotation procedure. 
The IP and OOP polarized emission rates, along with their sum, are shown for two emission polar angles, $\theta=60\degree$ and $\theta=0\degree$. 
The IP polarized emission is stronger than the OOP at $\theta=60\degree$, leading to a total PL that is anisotropic and maximal in the in-plane direction normal to the incident polarization. 
For $\theta=0$ (i.e., in the layer-normal direction) the two contributions are equal in magnitude and the resulting PL is isotropic. Both the IP and OOP polarizations lie in the $xy$ plane in the $\theta\rightarrow0$ limit, and the emitted photons are polarized in the $\hat{\mathbf{E}}_{\textrm{inc}}$ direction.\\ 
\indent
As seen from Eqs.~\ref{IPOOP}$-$\textcolor{blue}{7}, the OOP and IP radiative rates and PL signals are rotated by $\varphi = \pi/2$ with respect to one another, and their ratio is:
\begin{equation}
\label{ratio}
\frac{\gamma_S^{\rm OOP}(\varphi + \pi/2)}{\gamma_S^{\rm IP}(\varphi)} = \cos^2(\theta) \le 1.
\end{equation} 
This result explains why recent experiments \cite{Jones} observe a stronger PL signal polarized in plane compared to out of plane. 
When the linear polarization direction of the light that excites the sample is rotated (not shown), we find that only the total phase of the exciton wavefunction changes, and the PL pattern 
in Fig. \ref{fig2}(c) is unchanged but reoriented according to the linear polarization direction, in agreement with the measurements in Ref.~\cite{Jones}.\\
\indent 
%
% POLARIZER, JONES MATRICES
%
There is an important subtlety in the interpretation of recent PL measurements \cite{Wang, Ye-Heinz, Jones}. Due to the small size of the samples, the PL is typically collected through a microscope, 
measured in the layer-normal direction, and then passed through a polarizer or analyzer \cite{Wang, Ye-Heinz}. 
The resulting polar plots of the PL as a function of the angle $\alpha$ between the polarizer and the incident polarization exhibit a $\cos(2 \alpha)$ trend \cite{Jones, Wang, Ye-Heinz}. 
In these works, we feel that the dependence of the PL on the polarizer angle $\alpha$ has not been clearly differentiated from the PL dependence on emission direction. 
We stress that the PL anisotropy computed as a function of emission angle $\varphi$ in Fig.~\ref{fig2}(b,c) is \textit{distinct} from the PL anisotropy measured as a function polarizer angle $\alpha$, which can be readily explained with our approach.\\
\indent
In the $\theta \!\rightarrow\! 0$ limit probed experimentally, the radiative rate in Eq.~\ref{FGR} is $\gamma_S \!\propto\! \sum_{\lambda} \left|\bfe_{\l\bfq } \cdot \bfp_{S}\right|^2$. For excitation with polarization along $\hat{\mathbf{x}}$, which induces a dipole $\bfp_{S} = p_S \hat{\mathbf{x}}$, collecting light through a polarizer oriented at angle $\alpha$ gives $\gamma_S \propto p_S^2 \sum_{\lambda} \left|(A_\alpha \bfe_{\l\bfq }) \cdot \hat{\mathbf{x}}\right|^2$, where $A_\alpha$ is the \mbox{Jones matrix \cite{Jones-matrix}} 
\begin{equation}
A_\alpha = \begin{pmatrix} \cos^2\alpha & \cos\alpha\, \sin\alpha \\  \cos\alpha\, \sin\alpha & \sin^2\alpha \end{pmatrix}%\,.
\end{equation}
For $\theta \!\rightarrow\! 0$, one obtains easily $\gamma_S(\alpha) \propto p_S^2 \cos^2\alpha$, a result that also holds for arbitrary $\theta$. 
As a consequence, we predict a PL intensity as a function of polarizer angle $I(\alpha) = I_0\, \cos^2\alpha = I_0 [1 + \cos(2\alpha)] / 2$ (see Fig.~\ref{fig3}), which explains the $\cos(2 \alpha)$ angular dependence observed in the PL measurements \cite{Wang, Ye-Heinz, Jones}.\\
\indent 
Also shown in Fig.~\ref{fig3} is the expected PL intensity including exciton decoherence effects, which has a trend of $I(\alpha) = A_1 + A_2 \cos(2 \alpha)$ ($A_{i}$ are numerical constants). 
Two mechanisms can induce exciton decoherence, including $T_1$ relaxation processes, in which the exciton weights on the K and K' valleys vary due to intervalley scattering, resulting in exciton wavefunctions similar to Fig.~\ref{fig2}(b), and $T_2$ relaxation processes, in which the valley weights remain equal, but the exciton dipole $-$ and thus the polarization $-$ rotates by a random angle. 
Decoherence due to both processes opens a neck opens in the $I(\alpha)$ PL polar plot (see Fig.~\ref{fig3}) since a polarizer placed normal to the incident polarization will measure a non-zero signal. 
Recent measurements of $T_2$ times of $\sim$350 fs \cite{Wang,Ye-Heinz} at low temperature, where the radiative lifetime is of order 1$-$10 ps \cite{Palummo}, 
justify the significant loss of polarization observed experimentally \cite{Wang, Ye-Heinz}.\\ 
\indent
Microscopically, exciton dynamics between formation and radiative recombination is intricate. While we treated the bright $A$ 1$s$ exciton as two-fold degenerate, 
recent work has shown that two exciton branches with a very small energy difference ($\sim$1 meV in MoS$_2$) are present at the light cone due to the exchange interaction \cite{Diana}.  
These exciton branches correspond to a particular basis in the nearly degenerate pseudospin space. In our notation, excitons in the lower branch with parabolic dispersion couple only to IP polarized light, and excitons in the upper branch with $v$-shaped dispersion only to OOP polarized light \cite{Diana}. Our approach, which treats these branches as degenerate, forms a single exciton $\ket{S_1}$ that contributes to both IP and OOP polarized emission, which is equivalent to summing over the nearly degenerate branches in Ref.~\cite{Diana}.\\
\indent
On this basis, mechanisms that scatter excitons between the two nearly degenerate branches result in loss of exciton polarization, while mechanisms leading to change in exciton momentum $\bfQ$ within the same branch enable emission at all angles while keeping the polarization fixed. The vast amount of experimental data \cite{Jones, Wang, Wang-inplane,Schmidt} showing that the incident polarization is partially retained in the PL, while light is emitted in all directions \cite{Wang,Wang-inplane}, lead us to speculate that intra-branch exciton scattering is faster than inter-branch at low temperature, 
likely due to scattering with defects that rapidly re-orients the exciton momentum. 
These conditions are essential to observe the anisotropic PL we predict at $\theta \ne 0$ in Fig.~\ref{fig2}(c), where excitation with linearly polarized light yields a PL with maximal intensity in the in-plane direction \textit{normal} to the exciton dipole (as in classical dipole radiation) rather than parallel to the exciton dipole as in the $I(\alpha)$ plots. To our knowledge, such direction dependent measurements have not yet been carried out.\\ 
%
%%% ----- FIGURE 3  
%
\begin{figure}[!t]
\includegraphics[scale=0.2]{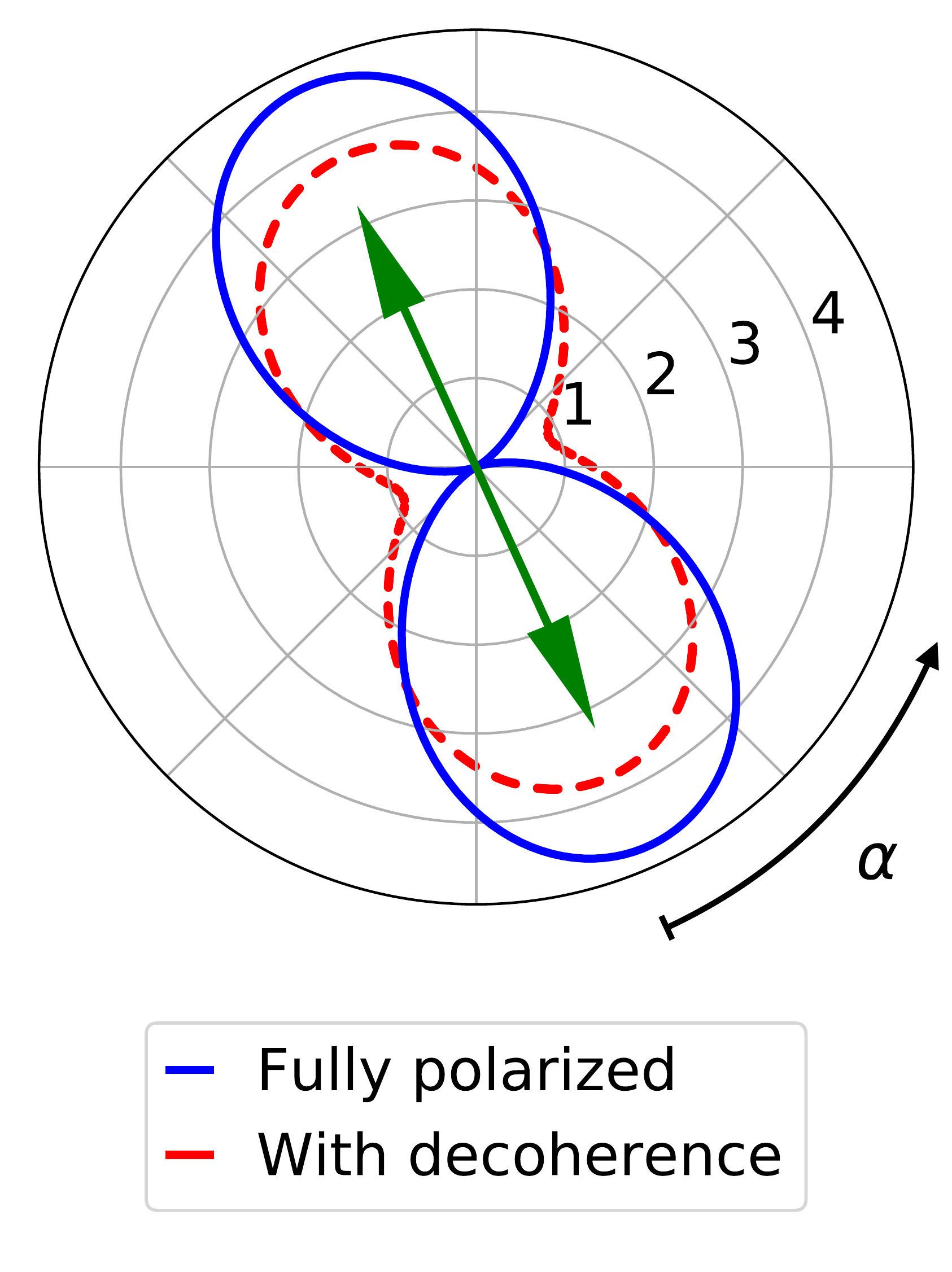}
\caption{Polar plot of the PL as a function of the angle $\alpha$ between the polarizer and the incident polarization. 
Shown are the ideal case in which light is fully polarized along the excitation polarization direction (indicated by the arrow) and the case in which light is only partially polarized as a result of decoherence.}
\label{fig3}
\end{figure} 
%
% RECOVER EQUATIONS IN OUR NANO LETTERS AS A LIMIT CASE
%
\indent
Lastly, we stress that our treatment generalizes the radiative rates derived in our previous work \cite{Palummo} under the assumption of isotropic exciton dipoles. %(not rigorously justified) 
When $\mathbf{p}_S$ is real and oriented along the $x \!=\! y$ direction, so that $p_x$ and $p_y$ are equal, Eq.~\ref{gamma0} reduces to our previously derived formula \cite{Palummo}, 
$\gamma_S(\mathbf{Q}) = \gamma_S(0) \cdot \sqrt{1 - \hbar^2c^2Q^2/E^2_S(\mathbf{Q})}$ \footnote{We remark that $\gamma_S(0)=\frac{e^2p_S^2}{\epsilon_0m^2cA E_S(0)}$ derived here is a factor of 2 smaller than in Ref. \cite{Palummo}, where the unit vector along the exciton dipole was taken to be $\hat{\bf x}+\hat{\bf y}$, and thus incorrectly normalized to $\sqrt{2}$ instead of 1. Note also that here we use SI units, whereas Ref.~\cite{Palummo} uses CGS units, in which $\epsilon_0 = 1 / 4\pi$, and further substitutes $p_S^2 = m^2 E^2_S(0) \mu^2_S / \hbar^2$.}. 
%
% THERMAL EFFECTS AND THERMAL AVERAGES
%
The temperature dependence of the radiative rates in Ref.~\cite{Palummo} can also be recovered within the treatment presented here. 
The results discussed so far neglect thermal effects, and assume that excitons with any momentum are available for light emission. 
Exciton decoherence due to Coulomb and electron-phonon interactions occurs on a ps timescale \cite{Alejandro}, which is comparable with the radiative lifetime (1$-$10 ps) at low temperature and faster than the radiative lifetime (1$-$10 ns) at room temperature \cite{Palummo}. 
As the temperature increases, excitons are thus expected to decay radiatively from a thermal equilibrium distribution over $Q$.   
To include thermal effects (see Supplemental Material~\cite{Supp}), we average the radiative rate in Eq.~\ref{gamma0} over momentum $Q$ and obtain temperature dependent radiative rates, which when averaged over the emission angle $\varphi$ give the temperature dependent radiative lifetime derived in our previous work \cite{Palummo}:
\begin{equation}
\<\tau_S\> (T)=\langle\gamma_S\rangle^{-1}
=
\gamma_S^{-1}(0)\cdot
\frac{3}{4}
\left(\frac{E_S(0)^2}{2M_Sc^2k_BT}\right)^{-1}.
\end{equation}
The few ps lifetimes at low temperature and few ns room temperature lifetimes we predicted with this formula \cite{Palummo} have now been confirmed by several experiments \cite{Shi,Korn,Lagarde,Libai,Tisdale}.\\
\indent
%
%%%%%      SUMMARY  /  CONCLUSION
%
In summary, we presented a general \textit{ab initio} method to compute the radiative rate and PL as a function of direction and polarization in 2D semiconductors.  
The new treatment reveals the inherently anisotropic PL of 2D-TMDCs and its dependence on polarization, valley occupation and decoherence. 
These results advance microscopic understanding of light emission in 2D-TMDCs.\\
\indent
H.-Y.C thanks the Taiwan Ministry of Education for fellowship support. M.B. acknowledges support by the National Science Foundation under Grant ACI-1642443, and partial support from the Space Solar Program Initiative at the California Institute of Technology. M.P. acknowledges financial support from the specific initiative NEMESYS of the Italian National Institute of Nuclear Physics (INFN) and EC for the RISE project CoExAN GA644076. D.S. acknowledges funding from the European Union project MaX Materials design at the eXascale H2020-EINFRA-2015-1, Grant Agreement No. 676598 and Nanoscience Foundries and Fine Analysis - Europe H2020-INFRAIA-2014-2015, Grant Agreement No. 654360. This research used resources of the National Energy Research Scientific Computing Center, a DOE Office of Science User Facility supported by the Office of Science of the U.S. Department of Energy under Contract No DE-AC02-05CH11231.
%
%\bibliography{TMD_ref}

\begin{thebibliography}{46}%
\makeatletter
\providecommand \@ifxundefined [1]{%
 \@ifx{#1\undefined}
}%
\providecommand \@ifnum [1]{%
 \ifnum #1\expandafter \@firstoftwo
 \else \expandafter \@secondoftwo
 \fi
}%
\providecommand \@ifx [1]{%
 \ifx #1\expandafter \@firstoftwo
 \else \expandafter \@secondoftwo
 \fi
}%
\providecommand \natexlab [1]{#1}%
\providecommand \enquote  [1]{``#1''}%
\providecommand \bibnamefont  [1]{#1}%
\providecommand \bibfnamefont [1]{#1}%
\providecommand \citenamefont [1]{#1}%
\providecommand \href@noop [0]{\@secondoftwo}%
\providecommand \href [0]{\begingroup \@sanitize@url \@href}%
\providecommand \@href[1]{\@@startlink{#1}\@@href}%
\providecommand \@@href[1]{\endgroup#1\@@endlink}%
\providecommand \@sanitize@url [0]{\catcode `\\12\catcode `\$12\catcode
  `\&12\catcode `\#12\catcode `\^12\catcode `\_12\catcode `\%12\relax}%
\providecommand \@@startlink[1]{}%
\providecommand \@@endlink[0]{}%
\providecommand \url  [0]{\begingroup\@sanitize@url \@url }%
\providecommand \@url [1]{\endgroup\@href {#1}{\urlprefix }}%
\providecommand \urlprefix  [0]{URL }%
\providecommand \Eprint [0]{\href }%
\providecommand \doibase [0]{http://dx.doi.org/}%
\providecommand \selectlanguage [0]{\@gobble}%
\providecommand \bibinfo  [0]{\@secondoftwo}%
\providecommand \bibfield  [0]{\@secondoftwo}%
\providecommand \translation [1]{[#1]}%
\providecommand \BibitemOpen [0]{}%
\providecommand \bibitemStop [0]{}%
\providecommand \bibitemNoStop [0]{.\EOS\space}%
\providecommand \EOS [0]{\spacefactor3000\relax}%
\providecommand \BibitemShut  [1]{\csname bibitem#1\endcsname}%
\let\auto@bib@innerbib\@empty
%</preamble>
\bibitem [{\citenamefont {Wang}\ \emph {et~al.}(2012)\citenamefont {Wang},
  \citenamefont {Kalantar-Zadeh}, \citenamefont {Kis}, \citenamefont
  {Coleman},\ and\ \citenamefont {Strano}}]{Strano}%
  \BibitemOpen
  \bibfield  {author} {\bibinfo {author} {\bibfnamefont {Q.~H.}\ \bibnamefont
  {Wang}}, \bibinfo {author} {\bibfnamefont {K.}~\bibnamefont
  {Kalantar-Zadeh}}, \bibinfo {author} {\bibfnamefont {A.}~\bibnamefont {Kis}},
  \bibinfo {author} {\bibfnamefont {J.~N.}\ \bibnamefont {Coleman}}, \ and\
  \bibinfo {author} {\bibfnamefont {M.~S.}\ \bibnamefont {Strano}},\ }\href
  {\doibase 10.1038/nnano.2012.193} {\bibfield  {journal} {\bibinfo  {journal}
  {Nat. Nanotech.}\ }\textbf {\bibinfo {volume} {7}},\ \bibinfo {pages} {699}
  (\bibinfo {year} {2012})}\BibitemShut {NoStop}%
\bibitem [{\citenamefont {Eda}\ and\ \citenamefont {Maier}(2013)}]{Maier}%
  \BibitemOpen
  \bibfield  {author} {\bibinfo {author} {\bibfnamefont {G.}~\bibnamefont
  {Eda}}\ and\ \bibinfo {author} {\bibfnamefont {S.~A.}\ \bibnamefont
  {Maier}},\ }\href {\doibase 10.1021/nn403159y} {\bibfield  {journal}
  {\bibinfo  {journal} {ACS Nano}\ }\textbf {\bibinfo {volume} {7}},\ \bibinfo
  {pages} {5660} (\bibinfo {year} {2013})}\BibitemShut {NoStop}%
\bibitem [{\citenamefont {Bernardi}\ \emph {et~al.}(2017)\citenamefont
  {Bernardi}, \citenamefont {Ataca}, \citenamefont {Palummo},\ and\
  \citenamefont {Grossman}}]{Bernardi-review}%
  \BibitemOpen
  \bibfield  {author} {\bibinfo {author} {\bibfnamefont {M.}~\bibnamefont
  {Bernardi}}, \bibinfo {author} {\bibfnamefont {C.}~\bibnamefont {Ataca}},
  \bibinfo {author} {\bibfnamefont {M.}~\bibnamefont {Palummo}}, \ and\
  \bibinfo {author} {\bibfnamefont {J.~C.}\ \bibnamefont {Grossman}},\ }\href
  {\doibase 10.1515/nanoph-2015-0030} {\bibfield  {journal} {\bibinfo
  {journal} {Nanophotonics}\ }\textbf {\bibinfo {volume} {6}},\ \bibinfo
  {pages} {479} (\bibinfo {year} {2017})}\BibitemShut {NoStop}%
\bibitem [{\citenamefont {Bernardi}\ \emph {et~al.}(2013)\citenamefont
  {Bernardi}, \citenamefont {Palummo},\ and\ \citenamefont
  {Grossman}}]{Bernardi-ML}%
  \BibitemOpen
  \bibfield  {author} {\bibinfo {author} {\bibfnamefont {M.}~\bibnamefont
  {Bernardi}}, \bibinfo {author} {\bibfnamefont {M.}~\bibnamefont {Palummo}}, \
  and\ \bibinfo {author} {\bibfnamefont {J.~C.}\ \bibnamefont {Grossman}},\
  }\href {\doibase 10.1021/nl401544y} {\bibfield  {journal} {\bibinfo
  {journal} {Nano Lett.}\ }\textbf {\bibinfo {volume} {13}},\ \bibinfo {pages}
  {3664} (\bibinfo {year} {2013})}\BibitemShut {NoStop}%
\bibitem [{\citenamefont {Britnell}\ \emph {et~al.}(2013)\citenamefont
  {Britnell}, \citenamefont {Ribeiro}, \citenamefont {Eckmann}, \citenamefont
  {Jalil}, \citenamefont {Belle}, \citenamefont {Mishchenko}, \citenamefont
  {Kim}, \citenamefont {Gorbachev}, \citenamefont {Georgiou}, \citenamefont
  {Morozov}, \citenamefont {Grigorenko}, \citenamefont {Geim}, \citenamefont
  {Casiraghi}, \citenamefont {Neto},\ and\ \citenamefont
  {Novoselov}}]{Britnell}%
  \BibitemOpen
  \bibfield  {author} {\bibinfo {author} {\bibfnamefont {L.}~\bibnamefont
  {Britnell}}, \bibinfo {author} {\bibfnamefont {R.~M.}\ \bibnamefont
  {Ribeiro}}, \bibinfo {author} {\bibfnamefont {A.}~\bibnamefont {Eckmann}},
  \bibinfo {author} {\bibfnamefont {R.}~\bibnamefont {Jalil}}, \bibinfo
  {author} {\bibfnamefont {B.~D.}\ \bibnamefont {Belle}}, \bibinfo {author}
  {\bibfnamefont {A.}~\bibnamefont {Mishchenko}}, \bibinfo {author}
  {\bibfnamefont {Y.-J.}\ \bibnamefont {Kim}}, \bibinfo {author} {\bibfnamefont
  {R.~V.}\ \bibnamefont {Gorbachev}}, \bibinfo {author} {\bibfnamefont
  {T.}~\bibnamefont {Georgiou}}, \bibinfo {author} {\bibfnamefont {S.~V.}\
  \bibnamefont {Morozov}}, \bibinfo {author} {\bibfnamefont {A.~N.}\
  \bibnamefont {Grigorenko}}, \bibinfo {author} {\bibfnamefont {A.~K.}\
  \bibnamefont {Geim}}, \bibinfo {author} {\bibfnamefont {C.}~\bibnamefont
  {Casiraghi}}, \bibinfo {author} {\bibfnamefont {A.~H.~C.}\ \bibnamefont
  {Neto}}, \ and\ \bibinfo {author} {\bibfnamefont {K.~S.}\ \bibnamefont
  {Novoselov}},\ }\href {\doibase 10.1126/science.1235547} {\bibfield
  {journal} {\bibinfo  {journal} {Science}\ }\textbf {\bibinfo {volume}
  {340}},\ \bibinfo {pages} {1311} (\bibinfo {year} {2013})}\BibitemShut
  {NoStop}%
\bibitem [{\citenamefont {Wong}\ \emph {et~al.}(2017)\citenamefont {Wong},
  \citenamefont {Jariwala}, \citenamefont {Tagliabue}, \citenamefont {Tat},
  \citenamefont {Davoyan}, \citenamefont {Sherrott},\ and\ \citenamefont
  {Atwater}}]{Atwater}%
  \BibitemOpen
  \bibfield  {author} {\bibinfo {author} {\bibfnamefont {J.}~\bibnamefont
  {Wong}}, \bibinfo {author} {\bibfnamefont {D.}~\bibnamefont {Jariwala}},
  \bibinfo {author} {\bibfnamefont {G.}~\bibnamefont {Tagliabue}}, \bibinfo
  {author} {\bibfnamefont {K.}~\bibnamefont {Tat}}, \bibinfo {author}
  {\bibfnamefont {A.~R.}\ \bibnamefont {Davoyan}}, \bibinfo {author}
  {\bibfnamefont {M.~C.}\ \bibnamefont {Sherrott}}, \ and\ \bibinfo {author}
  {\bibfnamefont {H.~A.}\ \bibnamefont {Atwater}},\ }\href {\doibase
  10.1021/acsnano.7b03148} {\bibfield  {journal} {\bibinfo  {journal} {ACS
  Nano}\ }\textbf {\bibinfo {volume} {11}},\ \bibinfo {pages} {7230} (\bibinfo
  {year} {2017})}\BibitemShut {NoStop}%
\bibitem [{\citenamefont {Splendiani}\ \emph {et~al.}(2010)\citenamefont
  {Splendiani}, \citenamefont {Sun}, \citenamefont {Zhang}, \citenamefont {Li},
  \citenamefont {Kim}, \citenamefont {Chim}, \citenamefont {Galli},\ and\
  \citenamefont {Wang}}]{Splendiani}%
  \BibitemOpen
  \bibfield  {author} {\bibinfo {author} {\bibfnamefont {A.}~\bibnamefont
  {Splendiani}}, \bibinfo {author} {\bibfnamefont {L.}~\bibnamefont {Sun}},
  \bibinfo {author} {\bibfnamefont {Y.}~\bibnamefont {Zhang}}, \bibinfo
  {author} {\bibfnamefont {T.}~\bibnamefont {Li}}, \bibinfo {author}
  {\bibfnamefont {J.}~\bibnamefont {Kim}}, \bibinfo {author} {\bibfnamefont
  {C.-Y.}\ \bibnamefont {Chim}}, \bibinfo {author} {\bibfnamefont
  {G.}~\bibnamefont {Galli}}, \ and\ \bibinfo {author} {\bibfnamefont
  {F.}~\bibnamefont {Wang}},\ }\href {\doibase 10.1021/nl903868w} {\bibfield
  {journal} {\bibinfo  {journal} {Nano Lett.}\ }\textbf {\bibinfo {volume}
  {10}},\ \bibinfo {pages} {1271} (\bibinfo {year} {2010})}\BibitemShut
  {NoStop}%
\bibitem [{\citenamefont {Mak}\ \emph {et~al.}(2010)\citenamefont {Mak},
  \citenamefont {Lee}, \citenamefont {Hone}, \citenamefont {Shan},\ and\
  \citenamefont {Heinz}}]{Heinz}%
  \BibitemOpen
  \bibfield  {author} {\bibinfo {author} {\bibfnamefont {K.~F.}\ \bibnamefont
  {Mak}}, \bibinfo {author} {\bibfnamefont {C.}~\bibnamefont {Lee}}, \bibinfo
  {author} {\bibfnamefont {J.}~\bibnamefont {Hone}}, \bibinfo {author}
  {\bibfnamefont {J.}~\bibnamefont {Shan}}, \ and\ \bibinfo {author}
  {\bibfnamefont {T.~F.}\ \bibnamefont {Heinz}},\ }\href {\doibase
  10.1103/PhysRevLett.105.136805} {\bibfield  {journal} {\bibinfo  {journal}
  {Phys. Rev. Lett.}\ }\textbf {\bibinfo {volume} {105}},\ \bibinfo {pages}
  {136805} (\bibinfo {year} {2010})}\BibitemShut {NoStop}%
\bibitem [{\citenamefont {Palummo}\ \emph {et~al.}(2015)\citenamefont
  {Palummo}, \citenamefont {Bernardi},\ and\ \citenamefont
  {Grossman}}]{Palummo}%
  \BibitemOpen
  \bibfield  {author} {\bibinfo {author} {\bibfnamefont {M.}~\bibnamefont
  {Palummo}}, \bibinfo {author} {\bibfnamefont {M.}~\bibnamefont {Bernardi}}, \
  and\ \bibinfo {author} {\bibfnamefont {J.~C.}\ \bibnamefont {Grossman}},\
  }\href {\doibase 10.1021/nl503799t} {\bibfield  {journal} {\bibinfo
  {journal} {Nano Lett.}\ }\textbf {\bibinfo {volume} {15}},\ \bibinfo {pages}
  {2794} (\bibinfo {year} {2015})}\BibitemShut {NoStop}%
\bibitem [{\citenamefont {Amani}\ \emph {et~al.}(2015)\citenamefont {Amani},
  \citenamefont {Lien}, \citenamefont {Kiriya}, \citenamefont {Xiao},
  \citenamefont {Azcatl}, \citenamefont {Noh}, \citenamefont {Madhvapathy},
  \citenamefont {Addou}, \citenamefont {KC}, \citenamefont {Dubey},
  \citenamefont {Cho}, \citenamefont {Wallace}, \citenamefont {Lee},
  \citenamefont {He}, \citenamefont {Ager}, \citenamefont {Zhang},
  \citenamefont {Yablonovitch},\ and\ \citenamefont {Javey}}]{Amani}%
  \BibitemOpen
  \bibfield  {author} {\bibinfo {author} {\bibfnamefont {M.}~\bibnamefont
  {Amani}}, \bibinfo {author} {\bibfnamefont {D.-H.}\ \bibnamefont {Lien}},
  \bibinfo {author} {\bibfnamefont {D.}~\bibnamefont {Kiriya}}, \bibinfo
  {author} {\bibfnamefont {J.}~\bibnamefont {Xiao}}, \bibinfo {author}
  {\bibfnamefont {A.}~\bibnamefont {Azcatl}}, \bibinfo {author} {\bibfnamefont
  {J.}~\bibnamefont {Noh}}, \bibinfo {author} {\bibfnamefont {S.~R.}\
  \bibnamefont {Madhvapathy}}, \bibinfo {author} {\bibfnamefont
  {R.}~\bibnamefont {Addou}}, \bibinfo {author} {\bibfnamefont
  {S.}~\bibnamefont {KC}}, \bibinfo {author} {\bibfnamefont {M.}~\bibnamefont
  {Dubey}}, \bibinfo {author} {\bibfnamefont {K.}~\bibnamefont {Cho}}, \bibinfo
  {author} {\bibfnamefont {R.~M.}\ \bibnamefont {Wallace}}, \bibinfo {author}
  {\bibfnamefont {S.-C.}\ \bibnamefont {Lee}}, \bibinfo {author} {\bibfnamefont
  {J.-H.}\ \bibnamefont {He}}, \bibinfo {author} {\bibfnamefont {J.~W.}\
  \bibnamefont {Ager}}, \bibinfo {author} {\bibfnamefont {X.}~\bibnamefont
  {Zhang}}, \bibinfo {author} {\bibfnamefont {E.}~\bibnamefont {Yablonovitch}},
  \ and\ \bibinfo {author} {\bibfnamefont {A.}~\bibnamefont {Javey}},\ }\href
  {\doibase 10.1126/science.aad2114} {\bibfield  {journal} {\bibinfo  {journal}
  {Science}\ }\textbf {\bibinfo {volume} {350}},\ \bibinfo {pages} {1065}
  (\bibinfo {year} {2015})}\BibitemShut {NoStop}%
\bibitem [{\citenamefont {Goodman}\ \emph {et~al.}(2017)\citenamefont
  {Goodman}, \citenamefont {Willard},\ and\ \citenamefont {Tisdale}}]{Tisdale}%
  \BibitemOpen
  \bibfield  {author} {\bibinfo {author} {\bibfnamefont {A.~J.}\ \bibnamefont
  {Goodman}}, \bibinfo {author} {\bibfnamefont {A.~P.}\ \bibnamefont
  {Willard}}, \ and\ \bibinfo {author} {\bibfnamefont {W.~A.}\ \bibnamefont
  {Tisdale}},\ }\href {\doibase 10.1103/PhysRevB.96.121404} {\bibfield
  {journal} {\bibinfo  {journal} {Phys. Rev. B}\ }\textbf {\bibinfo {volume}
  {96}},\ \bibinfo {pages} {121404} (\bibinfo {year} {2017})}\BibitemShut
  {NoStop}%
\bibitem [{\citenamefont {Shi}\ \emph {et~al.}(2013)\citenamefont {Shi},
  \citenamefont {Yan}, \citenamefont {Bertolazzi}, \citenamefont {Brivio},
  \citenamefont {Gao}, \citenamefont {Kis}, \citenamefont {Jena}, \citenamefont
  {Xing},\ and\ \citenamefont {Huang}}]{Shi}%
  \BibitemOpen
  \bibfield  {author} {\bibinfo {author} {\bibfnamefont {H.}~\bibnamefont
  {Shi}}, \bibinfo {author} {\bibfnamefont {R.}~\bibnamefont {Yan}}, \bibinfo
  {author} {\bibfnamefont {S.}~\bibnamefont {Bertolazzi}}, \bibinfo {author}
  {\bibfnamefont {J.}~\bibnamefont {Brivio}}, \bibinfo {author} {\bibfnamefont
  {B.}~\bibnamefont {Gao}}, \bibinfo {author} {\bibfnamefont {A.}~\bibnamefont
  {Kis}}, \bibinfo {author} {\bibfnamefont {D.}~\bibnamefont {Jena}}, \bibinfo
  {author} {\bibfnamefont {H.~G.}\ \bibnamefont {Xing}}, \ and\ \bibinfo
  {author} {\bibfnamefont {L.}~\bibnamefont {Huang}},\ }\href {\doibase
  10.1021/nn303973r} {\bibfield  {journal} {\bibinfo  {journal} {ACS Nano}\
  }\textbf {\bibinfo {volume} {7}},\ \bibinfo {pages} {1072} (\bibinfo {year}
  {2013})}\BibitemShut {NoStop}%
\bibitem [{\citenamefont {Korn}\ \emph {et~al.}(2011)\citenamefont {Korn},
  \citenamefont {Heydrich}, \citenamefont {Hirmer}, \citenamefont
  {Schmutzler},\ and\ \citenamefont {Sch{\"u}ller}}]{Korn}%
  \BibitemOpen
  \bibfield  {author} {\bibinfo {author} {\bibfnamefont {T.}~\bibnamefont
  {Korn}}, \bibinfo {author} {\bibfnamefont {S.}~\bibnamefont {Heydrich}},
  \bibinfo {author} {\bibfnamefont {M.}~\bibnamefont {Hirmer}}, \bibinfo
  {author} {\bibfnamefont {J.}~\bibnamefont {Schmutzler}}, \ and\ \bibinfo
  {author} {\bibfnamefont {C.}~\bibnamefont {Sch{\"u}ller}},\ }\href {\doibase
  10.1063/1.3636402} {\bibfield  {journal} {\bibinfo  {journal} {Appl. Phys.
  Lett.}\ }\textbf {\bibinfo {volume} {99}},\ \bibinfo {pages} {102109}
  (\bibinfo {year} {2011})}\BibitemShut {NoStop}%
\bibitem [{\citenamefont {Lagarde}\ \emph {et~al.}(2014)\citenamefont
  {Lagarde}, \citenamefont {Bouet}, \citenamefont {Marie}, \citenamefont {Zhu},
  \citenamefont {Liu}, \citenamefont {Amand}, \citenamefont {Tan},\ and\
  \citenamefont {Urbaszek}}]{Lagarde}%
  \BibitemOpen
  \bibfield  {author} {\bibinfo {author} {\bibfnamefont {D.}~\bibnamefont
  {Lagarde}}, \bibinfo {author} {\bibfnamefont {L.}~\bibnamefont {Bouet}},
  \bibinfo {author} {\bibfnamefont {X.}~\bibnamefont {Marie}}, \bibinfo
  {author} {\bibfnamefont {C.~R.}\ \bibnamefont {Zhu}}, \bibinfo {author}
  {\bibfnamefont {B.~L.}\ \bibnamefont {Liu}}, \bibinfo {author} {\bibfnamefont
  {T.}~\bibnamefont {Amand}}, \bibinfo {author} {\bibfnamefont {P.~H.}\
  \bibnamefont {Tan}}, \ and\ \bibinfo {author} {\bibfnamefont
  {B.}~\bibnamefont {Urbaszek}},\ }\href {\doibase
  10.1103/PhysRevLett.112.047401} {\bibfield  {journal} {\bibinfo  {journal}
  {Phys. Rev. Lett.}\ }\textbf {\bibinfo {volume} {112}},\ \bibinfo {pages}
  {047401} (\bibinfo {year} {2014})}\BibitemShut {NoStop}%
\bibitem [{\citenamefont {Yuan}\ \emph {et~al.}(2017)\citenamefont {Yuan},
  \citenamefont {Wang}, \citenamefont {Zhu}, \citenamefont {Zhou},\ and\
  \citenamefont {Huang}}]{Libai}%
  \BibitemOpen
  \bibfield  {author} {\bibinfo {author} {\bibfnamefont {L.}~\bibnamefont
  {Yuan}}, \bibinfo {author} {\bibfnamefont {T.}~\bibnamefont {Wang}}, \bibinfo
  {author} {\bibfnamefont {T.}~\bibnamefont {Zhu}}, \bibinfo {author}
  {\bibfnamefont {M.}~\bibnamefont {Zhou}}, \ and\ \bibinfo {author}
  {\bibfnamefont {L.}~\bibnamefont {Huang}},\ }\href {\doibase
  10.1021/acs.jpclett.7b00885} {\bibfield  {journal} {\bibinfo  {journal} {J.
  Phys. Chem. Lett.}\ }\textbf {\bibinfo {volume} {8}},\ \bibinfo {pages}
  {3371} (\bibinfo {year} {2017})}\BibitemShut {NoStop}%
\bibitem [{\citenamefont {Yao}\ \emph {et~al.}(2008)\citenamefont {Yao},
  \citenamefont {Xiao},\ and\ \citenamefont {Niu}}]{Xiao-2}%
  \BibitemOpen
  \bibfield  {author} {\bibinfo {author} {\bibfnamefont {W.}~\bibnamefont
  {Yao}}, \bibinfo {author} {\bibfnamefont {D.}~\bibnamefont {Xiao}}, \ and\
  \bibinfo {author} {\bibfnamefont {Q.}~\bibnamefont {Niu}},\ }\href {\doibase
  10.1103/PhysRevB.77.235406} {\bibfield  {journal} {\bibinfo  {journal} {Phys.
  Rev. B}\ }\textbf {\bibinfo {volume} {77}},\ \bibinfo {pages} {235406}
  (\bibinfo {year} {2008})}\BibitemShut {NoStop}%
\bibitem [{\citenamefont {Cao}\ \emph {et~al.}(2012)\citenamefont {Cao},
  \citenamefont {Wang}, \citenamefont {Han}, \citenamefont {Ye}, \citenamefont
  {Zhu}, \citenamefont {Shi}, \citenamefont {Niu}, \citenamefont {Tan},
  \citenamefont {Wang}, \citenamefont {Liu},\ and\ \citenamefont {Feng}}]{Cao}%
  \BibitemOpen
  \bibfield  {author} {\bibinfo {author} {\bibfnamefont {T.}~\bibnamefont
  {Cao}}, \bibinfo {author} {\bibfnamefont {G.}~\bibnamefont {Wang}}, \bibinfo
  {author} {\bibfnamefont {W.}~\bibnamefont {Han}}, \bibinfo {author}
  {\bibfnamefont {H.}~\bibnamefont {Ye}}, \bibinfo {author} {\bibfnamefont
  {C.}~\bibnamefont {Zhu}}, \bibinfo {author} {\bibfnamefont {J.}~\bibnamefont
  {Shi}}, \bibinfo {author} {\bibfnamefont {Q.}~\bibnamefont {Niu}}, \bibinfo
  {author} {\bibfnamefont {P.}~\bibnamefont {Tan}}, \bibinfo {author}
  {\bibfnamefont {E.}~\bibnamefont {Wang}}, \bibinfo {author} {\bibfnamefont
  {B.}~\bibnamefont {Liu}}, \ and\ \bibinfo {author} {\bibfnamefont
  {J.}~\bibnamefont {Feng}},\ }\href {http://dx.doi.org/10.1038/ncomms1882}
  {\bibfield  {journal} {\bibinfo  {journal} {Nat Commun.}\ }\textbf {\bibinfo
  {volume} {3}},\ \bibinfo {pages} {887} (\bibinfo {year} {2012})}\BibitemShut
  {NoStop}%
\bibitem [{\citenamefont {Xu}\ \emph {et~al.}(2014)\citenamefont {Xu},
  \citenamefont {Yao}, \citenamefont {Xiao},\ and\ \citenamefont
  {Heinz}}]{Xu-spin}%
  \BibitemOpen
  \bibfield  {author} {\bibinfo {author} {\bibfnamefont {X.}~\bibnamefont
  {Xu}}, \bibinfo {author} {\bibfnamefont {W.}~\bibnamefont {Yao}}, \bibinfo
  {author} {\bibfnamefont {D.}~\bibnamefont {Xiao}}, \ and\ \bibinfo {author}
  {\bibfnamefont {T.~F.}\ \bibnamefont {Heinz}},\ }\href {\doibase
  10.1038/nphys2942} {\bibfield  {journal} {\bibinfo  {journal} {Nat. Phys.}\
  }\textbf {\bibinfo {volume} {10}},\ \bibinfo {pages} {343} (\bibinfo {year}
  {2014})}\BibitemShut {NoStop}%
\bibitem [{\citenamefont {Zeng}\ \emph {et~al.}(2012)\citenamefont {Zeng},
  \citenamefont {Dai}, \citenamefont {Yao}, \citenamefont {Xiao},\ and\
  \citenamefont {Cui}}]{Zeng}%
  \BibitemOpen
  \bibfield  {author} {\bibinfo {author} {\bibfnamefont {H.}~\bibnamefont
  {Zeng}}, \bibinfo {author} {\bibfnamefont {J.}~\bibnamefont {Dai}}, \bibinfo
  {author} {\bibfnamefont {W.}~\bibnamefont {Yao}}, \bibinfo {author}
  {\bibfnamefont {D.}~\bibnamefont {Xiao}}, \ and\ \bibinfo {author}
  {\bibfnamefont {X.}~\bibnamefont {Cui}},\ }\href {\doibase
  10.1038/nnano.2012.95} {\bibfield  {journal} {\bibinfo  {journal} {Nat.
  Nanotech.}\ }\textbf {\bibinfo {volume} {7}},\ \bibinfo {pages} {490}
  (\bibinfo {year} {2012})}\BibitemShut {NoStop}%
\bibitem [{\citenamefont {Mak}\ \emph {et~al.}(2012)\citenamefont {Mak},
  \citenamefont {He}, \citenamefont {Shan},\ and\ \citenamefont
  {Heinz}}]{Mak-valley}%
  \BibitemOpen
  \bibfield  {author} {\bibinfo {author} {\bibfnamefont {K.~F.}\ \bibnamefont
  {Mak}}, \bibinfo {author} {\bibfnamefont {K.}~\bibnamefont {He}}, \bibinfo
  {author} {\bibfnamefont {J.}~\bibnamefont {Shan}}, \ and\ \bibinfo {author}
  {\bibfnamefont {T.~F.}\ \bibnamefont {Heinz}},\ }\href {\doibase
  10.1038/nnano.2012.96} {\bibfield  {journal} {\bibinfo  {journal} {Nat.
  Nanotech.}\ }\textbf {\bibinfo {volume} {7}},\ \bibinfo {pages} {494}
  (\bibinfo {year} {2012})}\BibitemShut {NoStop}%
\bibitem [{\citenamefont {Xiao}\ \emph {et~al.}(2012)\citenamefont {Xiao},
  \citenamefont {Liu}, \citenamefont {Feng}, \citenamefont {Xu},\ and\
  \citenamefont {Yao}}]{Xiao}%
  \BibitemOpen
  \bibfield  {author} {\bibinfo {author} {\bibfnamefont {D.}~\bibnamefont
  {Xiao}}, \bibinfo {author} {\bibfnamefont {G.-B.}\ \bibnamefont {Liu}},
  \bibinfo {author} {\bibfnamefont {W.}~\bibnamefont {Feng}}, \bibinfo {author}
  {\bibfnamefont {X.}~\bibnamefont {Xu}}, \ and\ \bibinfo {author}
  {\bibfnamefont {W.}~\bibnamefont {Yao}},\ }\href {\doibase
  10.1103/PhysRevLett.108.196802} {\bibfield  {journal} {\bibinfo  {journal}
  {Phys. Rev. Lett.}\ }\textbf {\bibinfo {volume} {108}},\ \bibinfo {pages}
  {196802} (\bibinfo {year} {2012})}\BibitemShut {NoStop}%
\bibitem [{\citenamefont {Jones}\ \emph {et~al.}(2013)\citenamefont {Jones},
  \citenamefont {Yu}, \citenamefont {Ghimire}, \citenamefont {Wu},
  \citenamefont {Aivazian}, \citenamefont {Ross}, \citenamefont {Zhao},
  \citenamefont {Yan}, \citenamefont {Mandrus}, \citenamefont {Xiao},
  \citenamefont {Yao},\ and\ \citenamefont {Xu}}]{Jones}%
  \BibitemOpen
  \bibfield  {author} {\bibinfo {author} {\bibfnamefont {A.~M.}\ \bibnamefont
  {Jones}}, \bibinfo {author} {\bibfnamefont {H.}~\bibnamefont {Yu}}, \bibinfo
  {author} {\bibfnamefont {N.~J.}\ \bibnamefont {Ghimire}}, \bibinfo {author}
  {\bibfnamefont {S.}~\bibnamefont {Wu}}, \bibinfo {author} {\bibfnamefont
  {G.}~\bibnamefont {Aivazian}}, \bibinfo {author} {\bibfnamefont {J.~S.}\
  \bibnamefont {Ross}}, \bibinfo {author} {\bibfnamefont {B.}~\bibnamefont
  {Zhao}}, \bibinfo {author} {\bibfnamefont {J.}~\bibnamefont {Yan}}, \bibinfo
  {author} {\bibfnamefont {D.~G.}\ \bibnamefont {Mandrus}}, \bibinfo {author}
  {\bibfnamefont {D.}~\bibnamefont {Xiao}}, \bibinfo {author} {\bibfnamefont
  {W.}~\bibnamefont {Yao}}, \ and\ \bibinfo {author} {\bibfnamefont
  {X.}~\bibnamefont {Xu}},\ }\href {http://dx.doi.org/10.1038/nnano.2013.151}
  {\bibfield  {journal} {\bibinfo  {journal} {Nat. Nanotech.}\ }\textbf
  {\bibinfo {volume} {8}},\ \bibinfo {pages} {634} (\bibinfo {year}
  {2013})}\BibitemShut {NoStop}%
\bibitem [{\citenamefont {Wang}\ \emph {et~al.}(2016)\citenamefont {Wang},
  \citenamefont {Marie}, \citenamefont {Liu}, \citenamefont {Amand},
  \citenamefont {Robert}, \citenamefont {Cadiz}, \citenamefont {Renucci},\ and\
  \citenamefont {Urbaszek}}]{Wang}%
  \BibitemOpen
  \bibfield  {author} {\bibinfo {author} {\bibfnamefont {G.}~\bibnamefont
  {Wang}}, \bibinfo {author} {\bibfnamefont {X.}~\bibnamefont {Marie}},
  \bibinfo {author} {\bibfnamefont {B.}~\bibnamefont {Liu}}, \bibinfo {author}
  {\bibfnamefont {T.}~\bibnamefont {Amand}}, \bibinfo {author} {\bibfnamefont
  {C.}~\bibnamefont {Robert}}, \bibinfo {author} {\bibfnamefont
  {F.}~\bibnamefont {Cadiz}}, \bibinfo {author} {\bibfnamefont
  {P.}~\bibnamefont {Renucci}}, \ and\ \bibinfo {author} {\bibfnamefont
  {B.}~\bibnamefont {Urbaszek}},\ }\href {\doibase
  10.1103/PhysRevLett.117.187401} {\bibfield  {journal} {\bibinfo  {journal}
  {Phys. Rev. Lett.}\ }\textbf {\bibinfo {volume} {117}},\ \bibinfo {pages}
  {187401} (\bibinfo {year} {2016})}\BibitemShut {NoStop}%
\bibitem [{\citenamefont {Ye}\ \emph {et~al.}(2017)\citenamefont {Ye},
  \citenamefont {Sun},\ and\ \citenamefont {Heinz}}]{Ye-Heinz}%
  \BibitemOpen
  \bibfield  {author} {\bibinfo {author} {\bibfnamefont {Z.}~\bibnamefont
  {Ye}}, \bibinfo {author} {\bibfnamefont {D.}~\bibnamefont {Sun}}, \ and\
  \bibinfo {author} {\bibfnamefont {T.~F.}\ \bibnamefont {Heinz}},\ }\href
  {https://www.nature.com/articles/nphys3891} {\bibfield  {journal} {\bibinfo
  {journal} {Nat. Phys.}\ }\textbf {\bibinfo {volume} {13}},\ \bibinfo {pages}
  {26} (\bibinfo {year} {2017})}\BibitemShut {NoStop}%
\bibitem [{\citenamefont {Yu}\ and\ \citenamefont {Wu}(2014)}]{Yu-valley}%
  \BibitemOpen
  \bibfield  {author} {\bibinfo {author} {\bibfnamefont {T.}~\bibnamefont
  {Yu}}\ and\ \bibinfo {author} {\bibfnamefont {M.~W.}\ \bibnamefont {Wu}},\
  }\href {\doibase 10.1103/PhysRevB.89.205303} {\bibfield  {journal} {\bibinfo
  {journal} {Phys. Rev. B}\ }\textbf {\bibinfo {volume} {89}},\ \bibinfo
  {pages} {205303} (\bibinfo {year} {2014})}\BibitemShut {NoStop}%
\bibitem [{\citenamefont {Yu}\ and\ \citenamefont {Wu}(2016)}]{Yu-Hall}%
  \BibitemOpen
  \bibfield  {author} {\bibinfo {author} {\bibfnamefont {T.}~\bibnamefont
  {Yu}}\ and\ \bibinfo {author} {\bibfnamefont {M.~W.}\ \bibnamefont {Wu}},\
  }\href {\doibase 10.1103/PhysRevB.93.045414} {\bibfield  {journal} {\bibinfo
  {journal} {Phys. Rev. B}\ }\textbf {\bibinfo {volume} {93}},\ \bibinfo
  {pages} {045414} (\bibinfo {year} {2016})}\BibitemShut {NoStop}%
\bibitem [{\citenamefont {Molina-Sanchez}\ \emph {et~al.}(2017)\citenamefont
  {Molina-Sanchez}, \citenamefont {Sangalli}, \citenamefont {Wirtz},\ and\
  \citenamefont {Marini}}]{Alejandro}%
  \BibitemOpen
  \bibfield  {author} {\bibinfo {author} {\bibfnamefont {A.}~\bibnamefont
  {Molina-Sanchez}}, \bibinfo {author} {\bibfnamefont {D.}~\bibnamefont
  {Sangalli}}, \bibinfo {author} {\bibfnamefont {L.}~\bibnamefont {Wirtz}}, \
  and\ \bibinfo {author} {\bibfnamefont {A.}~\bibnamefont {Marini}},\ }\href
  {\doibase 10.1021/acs.nanolett.7b00175} {\bibfield  {journal} {\bibinfo
  {journal} {Nano Lett.}\ }\textbf {\bibinfo {volume} {17}},\ \bibinfo {pages}
  {4549} (\bibinfo {year} {2017})}\BibitemShut {NoStop}%
\bibitem [{\citenamefont {Mai}\ \emph {et~al.}(2013)\citenamefont {Mai},
  \citenamefont {Barrette}, \citenamefont {Yu}, \citenamefont {Semenov},
  \citenamefont {Kim}, \citenamefont {Cao},\ and\ \citenamefont
  {Gundogdu}}]{Mai}%
  \BibitemOpen
  \bibfield  {author} {\bibinfo {author} {\bibfnamefont {C.}~\bibnamefont
  {Mai}}, \bibinfo {author} {\bibfnamefont {A.}~\bibnamefont {Barrette}},
  \bibinfo {author} {\bibfnamefont {Y.}~\bibnamefont {Yu}}, \bibinfo {author}
  {\bibfnamefont {Y.~G.}\ \bibnamefont {Semenov}}, \bibinfo {author}
  {\bibfnamefont {K.~W.}\ \bibnamefont {Kim}}, \bibinfo {author} {\bibfnamefont
  {L.}~\bibnamefont {Cao}}, \ and\ \bibinfo {author} {\bibfnamefont
  {K.}~\bibnamefont {Gundogdu}},\ }\href {\doibase 10.1021/nl403742j}
  {\bibfield  {journal} {\bibinfo  {journal} {Nano Lett.}\ }\textbf {\bibinfo
  {volume} {14}},\ \bibinfo {pages} {202} (\bibinfo {year} {2013})}\BibitemShut
  {NoStop}%
\bibitem [{\citenamefont {Hao}\ \emph {et~al.}(2016)\citenamefont {Hao},
  \citenamefont {Moody}, \citenamefont {Wu}, \citenamefont {Dass},
  \citenamefont {Xu}, \citenamefont {Chen}, \citenamefont {Sun}, \citenamefont
  {Li}, \citenamefont {Li}, \citenamefont {MacDonald},\ and\ \citenamefont
  {Li}}]{Hao}%
  \BibitemOpen
  \bibfield  {author} {\bibinfo {author} {\bibfnamefont {K.}~\bibnamefont
  {Hao}}, \bibinfo {author} {\bibfnamefont {G.}~\bibnamefont {Moody}}, \bibinfo
  {author} {\bibfnamefont {F.}~\bibnamefont {Wu}}, \bibinfo {author}
  {\bibfnamefont {C.~K.}\ \bibnamefont {Dass}}, \bibinfo {author}
  {\bibfnamefont {L.}~\bibnamefont {Xu}}, \bibinfo {author} {\bibfnamefont
  {C.-H.}\ \bibnamefont {Chen}}, \bibinfo {author} {\bibfnamefont
  {L.}~\bibnamefont {Sun}}, \bibinfo {author} {\bibfnamefont {M.-Y.}\
  \bibnamefont {Li}}, \bibinfo {author} {\bibfnamefont {L.-J.}\ \bibnamefont
  {Li}}, \bibinfo {author} {\bibfnamefont {A.~H.}\ \bibnamefont {MacDonald}}, \
  and\ \bibinfo {author} {\bibfnamefont {X.}~\bibnamefont {Li}},\ }\href
  {http://dx.doi.org/10.1038/nphys3674} {\bibfield  {journal} {\bibinfo
  {journal} {Nat. Phys.}\ }\textbf {\bibinfo {volume} {12}},\ \bibinfo {pages}
  {677} (\bibinfo {year} {2016})}\BibitemShut {NoStop}%
\bibitem [{\citenamefont {Perdew}\ \emph {et~al.}(1996)\citenamefont {Perdew},
  \citenamefont {Burke},\ and\ \citenamefont {Ernzerhof}}]{PBE}%
  \BibitemOpen
  \bibfield  {author} {\bibinfo {author} {\bibfnamefont {J.~P.}\ \bibnamefont
  {Perdew}}, \bibinfo {author} {\bibfnamefont {K.}~\bibnamefont {Burke}}, \
  and\ \bibinfo {author} {\bibfnamefont {M.}~\bibnamefont {Ernzerhof}},\ }\href
  {\doibase 10.1103/PhysRevLett.77.3865} {\bibfield  {journal} {\bibinfo
  {journal} {Phys. Rev. Lett.}\ }\textbf {\bibinfo {volume} {77}},\ \bibinfo
  {pages} {3865} (\bibinfo {year} {1996})}\BibitemShut {NoStop}%
\bibitem [{\citenamefont {Giannozzi}\ \emph {et~al.}(2009)\citenamefont
  {Giannozzi}, \citenamefont {Baroni}, \citenamefont {Bonini}, \citenamefont
  {Calandra}, \citenamefont {Car}, \citenamefont {Cavazzoni}, \citenamefont
  {Ceresoli}, \citenamefont {Chiarotti}, \citenamefont {Cococcioni},
  \citenamefont {Dabo}, \citenamefont {{Dal Corso}}, \citenamefont
  {de~Gironcoli}, \citenamefont {Fabris}, \citenamefont {Fratesi},
  \citenamefont {Gebauer}, \citenamefont {Gerstmann}, \citenamefont
  {Gougoussis}, \citenamefont {Kokalj}, \citenamefont {Lazzeri}, \citenamefont
  {Martin-Samos}, \citenamefont {Marzari}, \citenamefont {Mauri}, \citenamefont
  {Mazzarello}, \citenamefont {Paolini}, \citenamefont {Pasquarello},
  \citenamefont {Paulatto}, \citenamefont {Sbraccia}, \citenamefont {Scandolo},
  \citenamefont {Sclauzero}, \citenamefont {Seitsonen}, \citenamefont
  {Smogunov}, \citenamefont {Umari},\ and\ \citenamefont {Wentzcovitch}}]{QE}%
  \BibitemOpen
  \bibfield  {author} {\bibinfo {author} {\bibfnamefont {P.}~\bibnamefont
  {Giannozzi}}, \bibinfo {author} {\bibfnamefont {S.}~\bibnamefont {Baroni}},
  \bibinfo {author} {\bibfnamefont {N.}~\bibnamefont {Bonini}}, \bibinfo
  {author} {\bibfnamefont {M.}~\bibnamefont {Calandra}}, \bibinfo {author}
  {\bibfnamefont {R.}~\bibnamefont {Car}}, \bibinfo {author} {\bibfnamefont
  {C.}~\bibnamefont {Cavazzoni}}, \bibinfo {author} {\bibfnamefont
  {D.}~\bibnamefont {Ceresoli}}, \bibinfo {author} {\bibfnamefont {G.~L.}\
  \bibnamefont {Chiarotti}}, \bibinfo {author} {\bibfnamefont {M.}~\bibnamefont
  {Cococcioni}}, \bibinfo {author} {\bibfnamefont {I.}~\bibnamefont {Dabo}},
  \bibinfo {author} {\bibfnamefont {A.}~\bibnamefont {{Dal Corso}}}, \bibinfo
  {author} {\bibfnamefont {S.}~\bibnamefont {de~Gironcoli}}, \bibinfo {author}
  {\bibfnamefont {S.}~\bibnamefont {Fabris}}, \bibinfo {author} {\bibfnamefont
  {G.}~\bibnamefont {Fratesi}}, \bibinfo {author} {\bibfnamefont
  {R.}~\bibnamefont {Gebauer}}, \bibinfo {author} {\bibfnamefont
  {U.}~\bibnamefont {Gerstmann}}, \bibinfo {author} {\bibfnamefont
  {C.}~\bibnamefont {Gougoussis}}, \bibinfo {author} {\bibfnamefont
  {A.}~\bibnamefont {Kokalj}}, \bibinfo {author} {\bibfnamefont
  {M.}~\bibnamefont {Lazzeri}}, \bibinfo {author} {\bibfnamefont
  {L.}~\bibnamefont {Martin-Samos}}, \bibinfo {author} {\bibfnamefont
  {N.}~\bibnamefont {Marzari}}, \bibinfo {author} {\bibfnamefont
  {F.}~\bibnamefont {Mauri}}, \bibinfo {author} {\bibfnamefont
  {R.}~\bibnamefont {Mazzarello}}, \bibinfo {author} {\bibfnamefont
  {S.}~\bibnamefont {Paolini}}, \bibinfo {author} {\bibfnamefont
  {A.}~\bibnamefont {Pasquarello}}, \bibinfo {author} {\bibfnamefont
  {L.}~\bibnamefont {Paulatto}}, \bibinfo {author} {\bibfnamefont
  {C.}~\bibnamefont {Sbraccia}}, \bibinfo {author} {\bibfnamefont
  {S.}~\bibnamefont {Scandolo}}, \bibinfo {author} {\bibfnamefont
  {G.}~\bibnamefont {Sclauzero}}, \bibinfo {author} {\bibfnamefont {A.~P.}\
  \bibnamefont {Seitsonen}}, \bibinfo {author} {\bibfnamefont {A.}~\bibnamefont
  {Smogunov}}, \bibinfo {author} {\bibfnamefont {P.}~\bibnamefont {Umari}}, \
  and\ \bibinfo {author} {\bibfnamefont {R.~M.}\ \bibnamefont {Wentzcovitch}},\
  }\href
  {http://iopscience.iop.org/article/10.1088/0953-8984/21/39/395502/meta}
  {\bibfield  {journal} {\bibinfo  {journal} {J. Phys. Condens. Matter}\
  }\textbf {\bibinfo {volume} {21}},\ \bibinfo {pages} {395502} (\bibinfo
  {year} {2009})}\BibitemShut {NoStop}%
\bibitem [{\citenamefont {Cheiwchanchamnangij}\ and\ \citenamefont
  {Lambrecht}(2012)}]{Lambrecht}%
  \BibitemOpen
  \bibfield  {author} {\bibinfo {author} {\bibfnamefont {T.}~\bibnamefont
  {Cheiwchanchamnangij}}\ and\ \bibinfo {author} {\bibfnamefont {W.~R.~L.}\
  \bibnamefont {Lambrecht}},\ }\href {\doibase 10.1103/PhysRevB.85.205302}
  {\bibfield  {journal} {\bibinfo  {journal} {Phys. Rev. B}\ }\textbf {\bibinfo
  {volume} {85}},\ \bibinfo {pages} {205302} (\bibinfo {year}
  {2012})}\BibitemShut {NoStop}%
\bibitem [{\citenamefont {Marini}\ \emph {et~al.}(2009)\citenamefont {Marini},
  \citenamefont {Hogan}, \citenamefont {Gr{\"u}ning},\ and\ \citenamefont
  {Varsano}}]{Yambo}%
  \BibitemOpen
  \bibfield  {author} {\bibinfo {author} {\bibfnamefont {A.}~\bibnamefont
  {Marini}}, \bibinfo {author} {\bibfnamefont {C.}~\bibnamefont {Hogan}},
  \bibinfo {author} {\bibfnamefont {M.}~\bibnamefont {Gr{\"u}ning}}, \ and\
  \bibinfo {author} {\bibfnamefont {D.}~\bibnamefont {Varsano}},\ }\href
  {\doibase https://doi.org/10.1016/j.cpc.2009.02.003} {\bibfield  {journal}
  {\bibinfo  {journal} {Comput. Phys. Commun.}\ }\textbf {\bibinfo {volume}
  {180}},\ \bibinfo {pages} {1392 } (\bibinfo {year} {2009})}\BibitemShut
  {NoStop}%
\bibitem [{\citenamefont {Loudon}(2000)}]{Loudon}%
  \BibitemOpen
  \bibfield  {author} {\bibinfo {author} {\bibfnamefont {R.}~\bibnamefont
  {Loudon}},\ }\href@noop {} {\emph {\bibinfo {title} {The Quantum Theory of
  Light}}}\ (\bibinfo  {publisher} {OUP Oxford},\ \bibinfo {year}
  {2000})\BibitemShut {NoStop}%
\bibitem [{Note1()}]{Note1}%
  \BibitemOpen
  \bibinfo {note} {In practice, we use the velocity operator, and compute the
  transition dipole as $\protect \mathbf {p}_S (\protect \mathbf {Q})\protect
  \tmspace -\thinmuskip {.1667em}=\protect \tmspace -\thinmuskip {.1667em} (-i
  m / \hbar ) \mathinner {\delimiter "426830A {G| \left [ \protect \mathbf
  {x},H \right ]| S\protect \mathbf {Q}}\delimiter "526930B }$ to correctly
  include the non-local part of the Hamiltonian \cite {Davide}.}\BibitemShut
  {Stop}%
\bibitem [{\citenamefont {Gatti}\ and\ \citenamefont
  {Sottile}(2013)}]{Gatti-BSE}%
  \BibitemOpen
  \bibfield  {author} {\bibinfo {author} {\bibfnamefont {M.}~\bibnamefont
  {Gatti}}\ and\ \bibinfo {author} {\bibfnamefont {F.}~\bibnamefont
  {Sottile}},\ }\href {\doibase 10.1103/PhysRevB.88.155113} {\bibfield
  {journal} {\bibinfo  {journal} {Phys. Rev. B}\ }\textbf {\bibinfo {volume}
  {88}},\ \bibinfo {pages} {155113} (\bibinfo {year} {2013})}\BibitemShut
  {NoStop}%
\bibitem [{\citenamefont {Qiu}\ \emph {et~al.}(2015)\citenamefont {Qiu},
  \citenamefont {Cao},\ and\ \citenamefont {Louie}}]{Diana}%
  \BibitemOpen
  \bibfield  {author} {\bibinfo {author} {\bibfnamefont {D.~Y.}\ \bibnamefont
  {Qiu}}, \bibinfo {author} {\bibfnamefont {T.}~\bibnamefont {Cao}}, \ and\
  \bibinfo {author} {\bibfnamefont {S.~G.}\ \bibnamefont {Louie}},\ }\href
  {\doibase 10.1103/PhysRevLett.115.176801} {\bibfield  {journal} {\bibinfo
  {journal} {Phys. Rev. Lett.}\ }\textbf {\bibinfo {volume} {115}},\ \bibinfo
  {pages} {176801} (\bibinfo {year} {2015})}\BibitemShut {NoStop}%
\bibitem [{Note2()}]{Note2}%
  \BibitemOpen
  \bibinfo {note} {The IP and OOP polarizations are also referred to in the
  literature as the horizontal and vertical polarizations, respectively, or the
  transverse (IP) and longitudinal (OOP) polarizations in Ref.~\cite
  {Diana}}\BibitemShut {NoStop}%
\bibitem [{Sup()}]{Supp}%
  \BibitemOpen
  \href@noop {} {}\bibinfo {note} {See Supplemental Material at [URL] for
  detailed derivations of Eq. (4), Eqs. (6$-$7), and Eq. (11).}\BibitemShut
  {Stop}%
\bibitem [{\citenamefont {Hall}(2015)}]{Lie}%
  \BibitemOpen
  \bibfield  {author} {\bibinfo {author} {\bibfnamefont {B.}~\bibnamefont
  {Hall}},\ }\href@noop {} {\emph {\bibinfo {title} {Lie Groups, Lie Algebras,
  and Representations: an Elementary Introduction}}}\ (\bibinfo  {publisher}
  {Springer},\ \bibinfo {year} {2015})\BibitemShut {NoStop}%
\bibitem [{\citenamefont {Rohlfing}\ and\ \citenamefont
  {Louie}(2000)}]{Rohlfing}%
  \BibitemOpen
  \bibfield  {author} {\bibinfo {author} {\bibfnamefont {M.}~\bibnamefont
  {Rohlfing}}\ and\ \bibinfo {author} {\bibfnamefont {S.~G.}\ \bibnamefont
  {Louie}},\ }\href {\doibase 10.1103/PhysRevB.62.4927} {\bibfield  {journal}
  {\bibinfo  {journal} {Phys. Rev. B}\ }\textbf {\bibinfo {volume} {62}},\
  \bibinfo {pages} {4927} (\bibinfo {year} {2000})}\BibitemShut {NoStop}%
\bibitem [{\citenamefont {Jones}(1941)}]{Jones-matrix}%
  \BibitemOpen
  \bibfield  {author} {\bibinfo {author} {\bibfnamefont {R.~C.}\ \bibnamefont
  {Jones}},\ }\href {\doibase 10.1364/JOSA.31.000488} {\bibfield  {journal}
  {\bibinfo  {journal} {J. Opt. Soc. Am.}\ }\textbf {\bibinfo {volume} {31}},\
  \bibinfo {pages} {488} (\bibinfo {year} {1941})}\BibitemShut {NoStop}%
\bibitem [{\citenamefont {Wang}\ \emph {et~al.}(2017)\citenamefont {Wang},
  \citenamefont {Robert}, \citenamefont {Glazov}, \citenamefont {Cadiz},
  \citenamefont {Courtade}, \citenamefont {Amand}, \citenamefont {Lagarde},
  \citenamefont {Taniguchi}, \citenamefont {Watanabe}, \citenamefont
  {Urbaszek},\ and\ \citenamefont {Marie}}]{Wang-inplane}%
  \BibitemOpen
  \bibfield  {author} {\bibinfo {author} {\bibfnamefont {G.}~\bibnamefont
  {Wang}}, \bibinfo {author} {\bibfnamefont {C.}~\bibnamefont {Robert}},
  \bibinfo {author} {\bibfnamefont {M.~M.}\ \bibnamefont {Glazov}}, \bibinfo
  {author} {\bibfnamefont {F.}~\bibnamefont {Cadiz}}, \bibinfo {author}
  {\bibfnamefont {E.}~\bibnamefont {Courtade}}, \bibinfo {author}
  {\bibfnamefont {T.}~\bibnamefont {Amand}}, \bibinfo {author} {\bibfnamefont
  {D.}~\bibnamefont {Lagarde}}, \bibinfo {author} {\bibfnamefont
  {T.}~\bibnamefont {Taniguchi}}, \bibinfo {author} {\bibfnamefont
  {K.}~\bibnamefont {Watanabe}}, \bibinfo {author} {\bibfnamefont
  {B.}~\bibnamefont {Urbaszek}}, \ and\ \bibinfo {author} {\bibfnamefont
  {X.}~\bibnamefont {Marie}},\ }\href {\doibase 10.1103/PhysRevLett.119.047401}
  {\bibfield  {journal} {\bibinfo  {journal} {Phys. Rev. Lett.}\ }\textbf
  {\bibinfo {volume} {119}},\ \bibinfo {pages} {047401} (\bibinfo {year}
  {2017})}\BibitemShut {NoStop}%
\bibitem [{\citenamefont {Schmidt}\ \emph {et~al.}(2016)\citenamefont
  {Schmidt}, \citenamefont {Arora}, \citenamefont {Plechinger}, \citenamefont
  {Nagler}, \citenamefont {Granados~del \'Aguila}, \citenamefont {Ballottin},
  \citenamefont {Christianen}, \citenamefont {Michaelis~de Vasconcellos},
  \citenamefont {Sch\"uller}, \citenamefont {Korn},\ and\ \citenamefont
  {Bratschitsch}}]{Schmidt}%
  \BibitemOpen
  \bibfield  {author} {\bibinfo {author} {\bibfnamefont {R.}~\bibnamefont
  {Schmidt}}, \bibinfo {author} {\bibfnamefont {A.}~\bibnamefont {Arora}},
  \bibinfo {author} {\bibfnamefont {G.}~\bibnamefont {Plechinger}}, \bibinfo
  {author} {\bibfnamefont {P.}~\bibnamefont {Nagler}}, \bibinfo {author}
  {\bibfnamefont {A.}~\bibnamefont {Granados~del \'Aguila}}, \bibinfo {author}
  {\bibfnamefont {M.~V.}\ \bibnamefont {Ballottin}}, \bibinfo {author}
  {\bibfnamefont {P.~C.~M.}\ \bibnamefont {Christianen}}, \bibinfo {author}
  {\bibfnamefont {S.}~\bibnamefont {Michaelis~de Vasconcellos}}, \bibinfo
  {author} {\bibfnamefont {C.}~\bibnamefont {Sch\"uller}}, \bibinfo {author}
  {\bibfnamefont {T.}~\bibnamefont {Korn}}, \ and\ \bibinfo {author}
  {\bibfnamefont {R.}~\bibnamefont {Bratschitsch}},\ }\href {\doibase
  10.1103/PhysRevLett.117.077402} {\bibfield  {journal} {\bibinfo  {journal}
  {Phys. Rev. Lett.}\ }\textbf {\bibinfo {volume} {117}},\ \bibinfo {pages}
  {077402} (\bibinfo {year} {2016})}\BibitemShut {NoStop}%
\bibitem [{Note3()}]{Note3}%
  \BibitemOpen
  \bibinfo {note} {We remark that $\gamma _S(0)=\protect \frac
  {e^2p_S^2}{\epsilon _0m^2cA E_S(0)}$ derived here is a factor of 2 smaller
  than in Ref. \cite {Palummo}, where the unit vector along the exciton dipole
  was taken to be $\protect \mathaccentV {hat}05E{\protect \bf x}+\protect
  \mathaccentV {hat}05E{\protect \bf y}$, and thus incorrectly normalized to
  $\protect \sqrt {2}$ instead of 1. Note also that here we use SI units,
  whereas Ref.~\cite {Palummo} uses CGS units, in which $\epsilon _0 = 1 / 4\pi
  $, and further substitutes $p_S^2 = m^2 E^2_S(0) \mu ^2_S / \hbar
  ^2$.}\BibitemShut {Stop}%
\bibitem [{\citenamefont {Sangalli}\ \emph {et~al.}(2017)\citenamefont
  {Sangalli}, \citenamefont {Berger}, \citenamefont {Attaccalite},
  \citenamefont {Gr\"uning},\ and\ \citenamefont {Romaniello}}]{Davide}%
  \BibitemOpen
  \bibfield  {author} {\bibinfo {author} {\bibfnamefont {D.}~\bibnamefont
  {Sangalli}}, \bibinfo {author} {\bibfnamefont {J.~A.}\ \bibnamefont
  {Berger}}, \bibinfo {author} {\bibfnamefont {C.}~\bibnamefont {Attaccalite}},
  \bibinfo {author} {\bibfnamefont {M.}~\bibnamefont {Gr\"uning}}, \ and\
  \bibinfo {author} {\bibfnamefont {P.}~\bibnamefont {Romaniello}},\ }\href
  {\doibase 10.1103/PhysRevB.95.155203} {\bibfield  {journal} {\bibinfo
  {journal} {Phys. Rev. B}\ }\textbf {\bibinfo {volume} {95}},\ \bibinfo
  {pages} {155203} (\bibinfo {year} {2017})}\BibitemShut {NoStop}%
\end{thebibliography}
%
\end{document}